\documentclass[aps,pra,final,twocolumn,showpacs]{revtex4}

\usepackage{graphicx}
\usepackage{amsmath}
\usepackage{amsfonts}
\usepackage{hyperref}

\newcommand{\blue} {\ensuremath{^1\!S_0\,-\,^1\!P_1}}
\newcommand{\clock}{\ensuremath{^1\!S_0\,-\,^3\!P_0}}
\newcommand{\drain}{\ensuremath{^1\!S_0\,-\,^3\!P_1}}
\newcommand{\phiat}{\ensuremath{\varphi^{at}}}
\newcommand{\phik}[1]{\ensuremath{\varphi_{\scriptscriptstyle #1}}}
\newcommand{\DeltaFF}{\ensuremath{\Delta_{\scriptscriptstyle F,F'}}}
\newcommand{\mean}[1]{\ensuremath{\langle#1\rangle}}

\begin{document}

\title{Non-destructive measurement of the transition probability in a Sr optical lattice clock}

\author{J\'er\^ome Lodewyck, Philip G. Westergaard and Pierre Lemonde}
\affiliation{LNE-SYRTE, Observatoire de Paris, CNRS, UPMC~; 61 avenue de l'Observatoire, 75014 Paris, France}

\begin{abstract}
	We present the experimental demonstration of non-destructive probing of the \clock\ clock transition probability in an optical lattice clock with $^{87}$Sr atoms. It is based on the phase shift induced by the atoms on a weak off-resonant laser beam. The method we propose is a differential measurement of this phase shift on two modulation sidebands with opposite detuning with respect to the \blue\ transition, allowing a detection limited by the photon shot noise. We have measured an atomic population of $10^4$ atoms with a signal to noise ratio of 100 per cycle, while keeping more than 95\% of the atoms in the optical lattice with a depth of 0.1~mK. The method proves simple and robust enough to be operated as part of the whole clock setup. This detection scheme enables us to reuse atoms for subsequent clock state interrogations, dramatically reducing the loading time and thereby improving the clock frequency stability.
\end{abstract}

\pacs{06.30.Ft, 42.62.Fi, 37.10.Jk, 42.50.Nn}
\maketitle

	Optical lattice clocks with neutral atoms have recently experienced dramatic improvements in both frequency stability and accuracy~\cite{katori2008, opticalpumping, calcium, yb}, now surpassing microwave standards~\cite{Bize05} and comparable with optical clocks using single ions~\cite{Rosenband08}. Nonetheless, large improvements are still possible, especially in terms of frequency stability. The current best recorded stability is $2 \cdot 10^{-15} /\sqrt{\tau}$~\cite{calcium} with $\tau$ being the averaging time expressed in seconds. This is about one order of magnitude above the expected quantum limit. With a reasonable number of atoms larger than $10^4$, one could even anticipate a quantum limit in the $10^{-17} /\sqrt{\tau}$ range.

The excess noise is due to the Dick effect~\cite{santarellidick, 1464-4266-5-2-373}. The discountinuous interrogation of the atoms introduces a sampling of the interrogation laser frequency noise which is folded to low frequencies. To overcome this degradation and take full advantage of the high signal to noise ratio that is potentially achievable with a large number of neutral atoms, one should reduce the laser noise and/or minimize the dead time in the clock cycle. This latter possibility has been somewhat overlooked in existing lattice clocks. In present experiments the clock transition probability is detected by collecting the fluorescence photons scattered from an intense probe laser. Atoms are correspondingly heated and escape the trap during the detection. Most of each clock cycle is therefore spent capturing the atoms, leading to duty cycles of typically 10\%. By keeping the atoms in the lattice between clock cycles, the duty cycle could be increased up to 80\% or more, leading to a significant improvement of the clock frequency stability. This could be achieved with a non-destructive method to measure the transition probability. We demonstrate here such a detection scheme for an $^{87}$Sr lattice clock.

	The scheme is based on the measurement of the phase shift accumulated by a weak probe beam when passing through the atomic cloud. A method using the same physical effect was demonstrated with Cs atoms in~\cite{PhysRevA.71.043807,QNDPolzik}. For an atomic gas with states $|J,F,m_F \rangle$ and $|J',F',m'_F \rangle$, light with polarization state $q$ and wavelength $\lambda$ detuned by $\DeltaFF$ from the transition between the two states will experience a phase shift
\begin{eqnarray}
	\label{eq:phi}
	& & \phiat = \frac{3 \lambda^2 (2 J'+ 1)}{4\pi S} \hspace{-1.3em} \sum_{F,m_F, F',m'_F}
	\hspace{-1.3em} N_{m_F}(2F'+1)(2F+1) \\
	& & \times {\underbrace{\left(  \begin{array}{c c c}
	F' & 1 & F \\
	m'_F & q & -m_F
	\end{array} \right) }_{\textrm{Wigner 3\textit{j} symbol}}}^2
	{\underbrace{\left\{  \begin{array}{c c c}
	J & J' & 1 \\
	F' & F & I
	\end{array} \right\}}_{\textrm{Wigner 6\textit{j} symbol}}}^2
	\frac{(\Gamma/2) \DeltaFF}{\DeltaFF^2+(\Gamma/2)^2}\nonumber
\end{eqnarray}
where $N_{m_F}$ is the atomic population in the hyper-fine substate $|F,m_F \rangle$, $\Gamma$ is the natural linewidth of the transition and $S$ is the cross-section of the atomic cloud. With a Gaussian distributed laser beam and atomic cloud, averaging the phase shift over the transverse directions gives $S = 2\pi(r_0^2 + w^2/4)$ where $r_0$ is the cloud standard deviation and $w$ the $1/e^2$ radius of the laser beam.

\begin{figure}
\begin{center}
	\includegraphics[width=0.8\columnwidth]{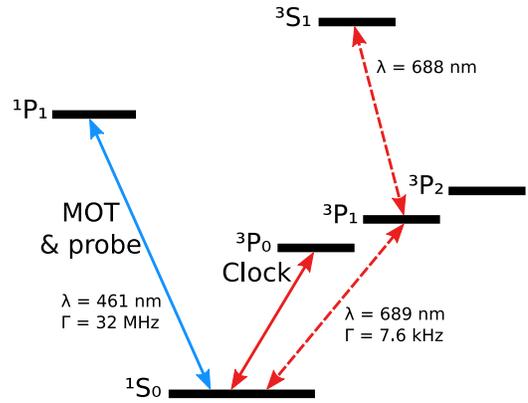}
	\caption{Energy levels of Sr of interest in this paper.}
	\label{fig:levels}
\end{center}
\end{figure}

\begin{figure}
\begin{center}
    \includegraphics[width=0.8\columnwidth]{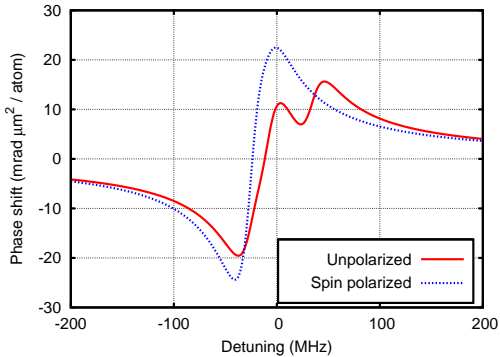}
    \caption{Theoretical phase shift $\phiat S/N$ for the \blue\ transition with zero magnetic field and a linearly polarized probe. It takes into account the three different $F' = 7/2,\ 9/2$ and $11/2$ levels of $^1\!P_1$, spanning over 60~MHz around their average frequency (center of the plot). The phase shift is represented for equally populated $m_F$ states (solid red curve) and spin-polarized atoms in $m_F = 9/2$ or $m_F = -9/2$ states (dashed blue curve). For a 90~MHz detuning, these phase shifts are comparable and amount to a few tens of mrad with typical parameters $N=10^4$~atoms and $S = 2.8\cdot 10^3 \mu$m$^2$.}
    \label{fig:signal}
\end{center}
\end{figure}

	To measure the clock transition probability, a suitable transition for the non-destructive scheme should be selected. There is no cycling transition involving the $^3P_0$ state so we considered only the \blue\ and \drain\ transitions at 461\,nm and 689\,nm respectively, which both involve the atomic ground state. At first sight, the \drain\ transition could seem more appealing due to the $\lambda^2$ dependence of $\phiat$. However, its small natural linewidth $\Gamma = 7.6$~kHz introduces experimental difficulties. Its Zeeman $m_F$ substates are resolved even for magnetic fields as low as 0.3~G, which would require to work at a large detuning thus dwarfing the phase signal. Further, the exact frequency of the probe would depend on the actual magnetic field and lattice induced light shift. For these reasons, we chose to operate with the more robust \blue\ transition, for which the induced phase shift is plotted in Fig.~\ref{fig:signal}.

\begin{figure}
\begin{center}
	\includegraphics[width=0.6\columnwidth]{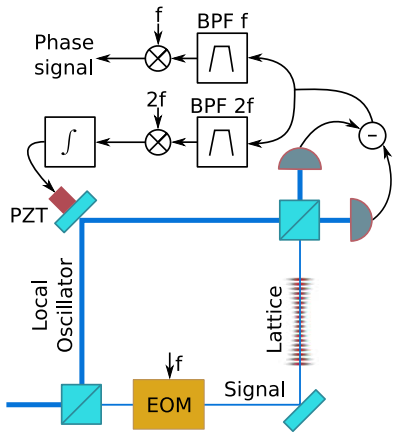}
	\caption{Experimental setup. The number of atoms in the optical lattice is proportionnal to the phase shift of the RF component at the modulation frequency $f$, filtered by a band-pass filter (BPF). The harmonic at frequency $2f$ is used to lock the phase of the interferometer, hence maximizing the RF power of the signal component.}
	\label{fig:setup}
\end{center}
\end{figure}

	We propose a phase shift measurement setup using an electro-optic phase modulator (EOM) in a Mach-Zender (MZ) interferometer (Fig.~\ref{fig:setup}). A laser beam resonant with the \blue\ transition is split into a weak signal (typically a few nW) and a strong local oscillator (LO) (a few mW). The signal beam is modulated at 90~MHz by the EOM before it is overlapped with the atoms in the optical lattice. The signal beam waist is $w = 37$~$\mu$m, comparable to the transverse size of the atomic ensemble ($r_0 = 10$~$\mu$m). The electric field of the signal beam is detected by a homodyne detection in which the signal interferes with the LO on a beam splitter and the light intensities in each output arm of the beam splitter are measured with fast Si photodiodes (Hamamatsu S5973) and electrically subtracted. In this scheme, the LO amplifies the signal without degrading its signal to noise ratio (SNR). For a LO power of 2~mW the electronic noise is smaller than the photon shot noise by a factor of 2.

	The RF output $s$ of the difference of the photocurrents is (up to a constant factor):
\begin{eqnarray}
	s = \sum_{n = 1}^{+\infty} J_n(a)g\left(\phi_0 - \frac{\phik{n} + \phik{-n}}{2}\right)g\left(n\omega t + \frac{\phik{n} - \phik{-n}}{2}\right) \label{eq:sRF}
\end{eqnarray}
	where $J_n$ is the Bessel function of the first kind, $\phi_0$ the phase of the LO, $a$ the modulation depth, $\omega$ the modulation angular frequency and $g = \cos$ ($\sin$) if $n$ is even (odd). $\phik{n}$ is the total phase shift experienced by the modulation sideband $n$. It can be expanded as $\phik{n} = \phiat(n\omega) + \delta\phik{n} + \phi_s$ where $\phiat(n\omega)$ is the atomic phase shift for a detuning $n\omega$ given by equation~(\ref{eq:phi}), $\phi_s$ is a global phase and $\delta\phik{n}$ is the laser phase noise. Because all the modulation sidebands belong to the same spatial mode, $\phi_s$ is independent of $n$. Given the low power at which we operate and the small line-width of our laser ($<$~1~MHz), $\delta\phik{n}$ is dominated by shot noise, even though our interferometer features an optical path difference of about 2~m between the signal and the LO.

	We can see from equation~(\ref{eq:sRF}) that the phase of the RF component at angular frequency $n\omega$ is the differential atomic phase shift of the $-n$ and $+n$ modulation sidebands. Since $\phiat$ is approximately an odd function of the probe detuning, this phase shift is proportional to the number of atoms in the atomic ground state. Furthermore, it does not depend on the phase $\phi_0$ of the LO nor the global phase $\phi_s$ of the signal, making our system independent of mechanical and thermal fluctuations. These features are very welcome given the small phase shifts we want to detect. However, the amplitude of the RF components does depend on $\phi_0$, and will eventually cross zero as $\phi_0$ drifts. The parity of $g$ shows in particular that the odd RF sidebands have maximum power when the amplitude of the even sidebands is null. We use this feature to lock $\phi_0$: we demodulate the second order RF component at angular frequency $2\,\omega$ and servo-loop $\phi_0$ with a piezoelectric transducer (PZT) to keep the demodulation signal at zero. The lock bandwidth is 10~kHz, limited by the mechanical properties of the PZT. Finally, the atomic phase signal is extracted by demodulating the first order RF component, maximized by the lock (Fig.~\ref{fig:setup}). We emphasize that the noise of this phase signal does not depend on the noise of the PZT lock to first order, due to the quadrature detection.

	In our setup, we choose the modulation frequency $f = \omega/2\pi$ and amplitude $a$ to optimize the signal to noise ratio (SNR) of the detection scheme. The final SNR results from a trade-off between the phase component of the optical shot noise which decreases at larger optical powers, and the heating of the atomic cloud which increases with the optical power as long as the transition is not saturated. Therefore we have to determine the optimal $f$ and $a$ for a given heating of the atoms. The signal to noise ratio is

\begin{eqnarray}
	SNR & = & \frac{\phiat(+\omega) - \phiat(-\omega)}{\sqrt{\mean{\delta\phik{1}^2} + \mean{\delta\phik{-1}^2}}}\label{eq:SNR}\\
	\textrm{with} \quad \mean{\delta\phik{+1}^2} & = & \mean{\delta\phik{-1}^2} = \frac{h c}{4\lambda|J_1(a)|^2 \eta PT} \label{eq:noise}
\end{eqnarray}
	where $P$ is the total optical power seen by the atoms, $T$ is the probe time and $\eta$ the detection efficiency. The product $PT$ is linked to the number of photons $n_\gamma$ absorbed by each atom of the atomic ensemble, characterizing the fraction of the atoms lost during the non-destructive probing,
\begin{equation}
	n_\gamma = PT \frac{\Gamma}{2P_{sat}} \sum_{n = -\infty}^{+\infty}\frac{|J_n(a)|^2}{1+ 4(n\omega)^2/\Gamma^2}.\label{eq:ngamma}
\end{equation}
Here, $P_{sat} = 1.2$~$\mu$W is the saturation power averaged over the atomic cloud. Combining equations~(\ref{eq:SNR}-\ref{eq:ngamma}) gives an expression of the SNR as a function of $\omega$ and $a$. We find that the SNR increases with $\omega$ and becomes nearly constant after a few $\Gamma$. In our experimental setup, we chose $\omega = 2\pi\,90$~MHz $ \simeq 3 \Gamma$ for which the SNR is nearly optimal. For this frequency, the optimal $a$ computed from equations~(\ref{eq:SNR}-\ref{eq:ngamma}) is very close to the modulation amplitude for which the resonant carrier is completely suppressed ($a = 2.4$ rad). Furthermore, the SNR is very flat around this optimum so that  temperature control of the EOM is not required. For $a= 2.4$ rad, the $+1$ and $-1$ modulation sidebands have 53\% of the optical power. The remaining power distributed in the higher order sidebands contributes to the heating of the atoms but not to the signal. From the previous equations, we calculate that these higher order sidebands degrade the SNR by only 8\%.

	Finally, the contrast of our interferometer is 76\% (measured with a balanced MZ configuration) and 25\% additional optical losses appear between the atoms and the detection. These defects are attributed to the vacuum chamber windows and the optics of the lattice cavity that were not originally designed to operate at the probe wavelength. As a result the detection efficiency is $\eta = 43$\%.

\begin{figure}
\begin{center}
	\includegraphics[width=\columnwidth]{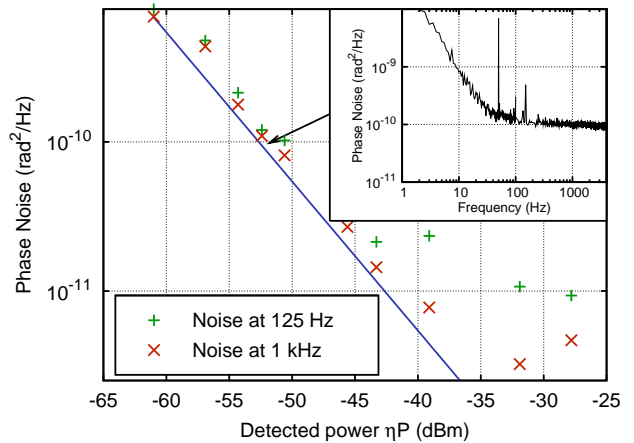}
	\caption{Detection noise ($\frac 1 2 \sqrt{\mean{\delta\phik{1}^2} + \mean{\delta\phik{-1}^2}}$) power spectral density at 125~Hz and 1~kHz. The signal is shot noise limited (blue line, from equation~(\ref{eq:noise})) for powers up to 30~nW. The inset shows the full phase noise spectrum for a typical detected optical power $\eta P = 5$~nW, corresponding to a white noise level of $10^{-10}$~rad$^2$/Hz.}
	\label{fig:fft}
\end{center}
\end{figure}

	We characterize the noise of the measurement setup as follows. A typical noise spectrum of the phase signal is shown in Fig.~\ref{fig:fft}. It is shot noise limited from 60~Hz. The detection system features a noise floor at $10^{-11}$~rad$^2$/Hz, so that the signal is shot noise limited for total optical powers up to 30~nW. Given this noise figure, we probe the atoms with $T=3$~ms signal pulses of typically $P = 12$~nW ($\eta P = 5$~nW). These pulses have a product $PT$ low enough to keep most of the atoms in the lattice (see below). They are short enough to escape the low frequency noise, and long compared to the PZT lock bandwidth.

	To measure the atomic population in $^1\!S_0$, we apply two consecutive probe pulses separated by a 7~ms interval. Between these pulses we shelve the atoms in the the dark states $^3\!P_0$ and $^3\!P_2$ by optical pumping on the $^1\!S_0\,-\,^3\!P_1$ and $^3\!P_1\,-\,^3\!S_1$ transitions. The second probe pulse does not experience the atomic phase shift and then acts as a zero phase reference. During the probe pulses, the phase signal is sampled at 500~kHz and the final signal is the difference of the averaged signal over each the probe duration. The noise of the resulting signal, as measured with no atoms in the lattice, is 0.4~mrad RMS for $\eta P = 5$~nW and scales as $1/\sqrt{P}$ as expected from equation~(\ref{eq:SNR}).

	With about $N =10^4$ atoms in the lattice, we measured a phase shift of 40~mrad corresponding to a SNR of 100 per cycle, which is close to the atomic shot noise.

	The measurement of the absolute transition probability associated with the interrogation of the atomic ensemble with our clock laser involves a third probe pulse. All probe pulses should be applied after the clock interrogation since low frequency phase drifts would add noise to the detection signal for long interrogation times. The sequence is as follows: after the clock interrogation a first probe pulse measures the number of atoms that remained in the atomic ground state. Then the atoms are repumped into the fundamental state, and are probed with a second probe pulse that determines the total number of atoms. Then, as before, we pump all the atoms into the dark states and apply a reference pulse. The measured noise on the transition probability is 2\% RMS with the previous parameters, and varies as $1/N$ for $N$ up to $10^4$.

\begin{figure}
\begin{center}
	\includegraphics[width=0.9\columnwidth]{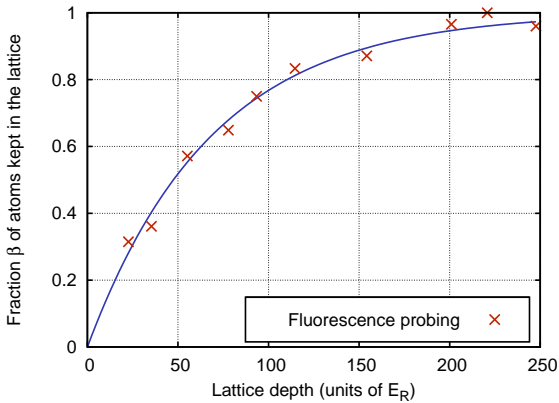}
	\caption{Atomic loss due to the probe laser for different lattice depths,
	measured by fluorescence detection after the non-destructive probing.
The solid curve is a fit of the fluorescence points with equation~(\ref{eq:beta}). This fit gives $n_\gamma$ = 103 photons per atom, in good agreement with the value of 81 photons per atom deduced from equation~(\ref{eq:ngamma}) and the measured optical power ($P = 14$~nW, $T = 3$~ms). For deep lattices, more than 95\% of the atoms remain trapped after the detection pulses.}
	\label{fig:heating}
\end{center}
\end{figure}
	A key feature of the detection scheme is the ability to recycle the atoms from one cycle to the other. To check that the detection pulses do not heat the atoms out of the lattice, we measured the atomic losses caused by the phase detection with a fluorescence probe at the end of each clock cycle. This measurement has been repeated for different lattice depths (Fig.~\ref{fig:heating}). We model the heating process by absorption and spontaneous emission. Because the probe beam is aligned with the lattice axis and the atoms are in the Lamb-Dicke regime in this direction, we assume that the recoil momentum associated with photon absorption is absorbed by the lattice and does not contribute to heating the atoms. However, the trapping potential is loose in the transverse directions, and therefore the horizontal component of the recoil momentum associated with spontaneous emission is entirely transferred to the atoms. For atoms initially in the vibrational ground state of the lattice, the fraction $\beta$ of atoms remaining in the lattice after detection is:
\begin{equation}
	\beta =  1 - \exp\left(-\frac{U_0/E_R}{2 n_\gamma/3}\right)\label{eq:beta}
\end{equation}
	where $U_0$ is the lattice depth, $E_R = \frac{h^2}{2m\lambda^2}$ is the recoil energy associated with the interaction between a Sr atom with mass $m$ and a probe photon ($\lambda = 461$~nm), and $n_\gamma$ is the number of absorbed photons per atom, given by equation~(\ref{eq:ngamma}). As shown in Fig.~\ref{fig:heating}, this model is in agreement with the experiment.

	We have experimentally demonstrated a non-destructive probing method for the transition probability in an optical lattice clock with Sr atoms. With a differential phase measurement of two modulation sidebands, we achieve a high detectivity without resorting to complex interferometric stabilization methods. This detectivity is intrinsically limited by the atomic transition we probe, and not by our detection system. We have integrated the measurement procedure in the clock cycle and demonstrated the feasibility of measuring the clock transition probability. By recycling the atoms we expect to be able to reduce the dead time of the clock cycle down to \emph{ca} 100 ms while keeping more than $10^3$ atoms in the experiment. Together with an improved clock laser currently under development~\cite{cavities}, this would open the way to better clock stabilities, below $10^{-15}/\sqrt{\tau}$. Furthermore, as recently shown~\cite{SSPolzik,vuletic}, the detection scheme presented here is capable of spin-squeezing the atomic ensemble, a technique that can ultimately overcome the atomic shot noise in atomic clocks.

We thank A. Clairon and K. Gibble for fruitful discussions. SYRTE is a member of IFRAF (Institut Francilien de Recherche sur les Atomes Froids). This work has received funding from the European Community's Seventh Framework Programme, ERA-NET Plus, under Grant Agreement No. 217257, as well as from IFRAF, CNES and ESA.

\bibliographystyle{apsrev}
\bibliography{qnd}

\end{document}